\documentclass[a4paper,11pt]{article}
\usepackage{jinstpub} 
\usepackage{lineno}
\usepackage{tikz}
\usetikzlibrary{calc}
\usepackage{subcaption}
\usepackage{siunitx}
\usepackage[english]{cleveref}
\newcommand{\fluenceUnits}[1]{\SI{#1}{n_{eq}\,\cm^{-2}}}

\DeclareSIUnit{\arbunit}{a.u.}

\title{\boldmath Tracking and timing measurements on irradiated TI-LGADs}

\author[a]{A. Gomez-Carrera\footnote{Corresponding author}}
\author[b]{L. Diehl}
\author[a]{J. Duarte-Campderros}
\author[a]{M. Fernandez}
\author[b]{V. Gkougkousis}
\author[c]{B. Hiti}
\author[c]{G. Kramberger}
\author[b]{A. Macchiolo}
\author[b]{Y. Padniuk}
\author[c]{M. Puklavec}
\author[d]{C. Torres Muñoz}
\author[c]{I. Velkovska}
\author[a]{I. Vila}

\affiliation[a]{Instituto de Física de Cantabria, IFCA (CSIC-UC)\\
Av. los Castros, Santander, Spain}       
\affiliation[b]{University of Zurich \\
Winterthurerstrasse 190, 8057 Zürich, Switzerland}
\affiliation[c]{Jozef Stefan Institute \\ 
Namova cesta 39, 1000 Ljubljana, Slovenia}
\affiliation[d]{Centro Nacional de Aceleradores \\
Calle Tomás Alba Edison, 7, Isla de Cartuja, 41092 Sevilla, Spain}

\emailAdd{antonio.gomez.carrera@cern.ch}

\abstract{Trench-isolated (TI) LGADs, developed at FBK, are pixelated LGAD implementations where pads are separated by physical trenches etched within the silicon substrate and filled with a dielectric material. Developed as a solution to the LGAD fill factor problem, this technology provides a path towards 4D tracking with reduced inefficiencies in the interpad regions. Through a dedicated 120 GeV SPS pion test beam campaign,
the time resolution, efficiency, and inter-pad distance of carbon infused irradiated TI-LGADs is presented. Fluences up to \fluenceUnits{2.5e15} are evaluated, for single trench implementations with varied trench width at a temperature of -25ºC. The results show an interpad distance degradation with the irradiation of the detector and an optimal time resolution between \SI{35}{\ps} and \SI{45}{\ps} for all studied devices.}

\keywords{Particle tracking detectors (Solid-state detectors); Radiation-hard detectors; Solid state
detectors; Timing detectors}

\arxivnumber{-} 

\begin{document}
\maketitle
\flushbottom

\section{Introduction}\label{sec:intro}

The High-Luminosity Large Hadron Collider (HL-LHC) aims to deliver an integrated luminosity of \SI{4000}{\per\femto\barn} over 10 years~\cite{aberle2020high}, creating up to 200 proton-proton interactions per bunch crossing. This high particle density complicates the association of reconstructed objects with the correct collision vertex. To mitigate this, MIP timing subdetectors are being developed, for the Phase-2 upgrades of ATLAS~\cite{ATLAS-TDR-MIPetector} and CMS~\cite{CMS-TDR-MIPDetector}, targeting a resolution of \SI{30}{\pico\s} to \SI{50}{\pico\s}.

\begin{figure}[htbp]
    \centering
    \includegraphics[width=0.4\linewidth]{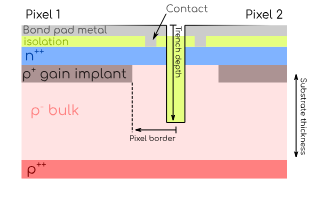}
    \caption{Transversal view of a Trench isolated Low Gain Avalanche Diode. The trench and multiplication layer are visible. Not to scale, source: \cite{senger2023TiLGAD}.}
    \label{fig Ti-LGAD scheme}
\end{figure}

Low Gain Avalanche Diodes (LGADs)~\cite{LGAD} technology has been selected for the endcap regions, due to their internal gain, fast response and radiation hardness which allow them to be closer to the beampipe than the SiPM selected for the barrel regions. However, their fill factor is limited by $\sim80~\si{\micro\metre}$ no-gain regions caused by edge termination structures. Trench-Isolated LGADs (TI-LGADs)~\cite{senger2023TiLGAD} address this by replacing junction terminations with dielectric-filled trenches, reducing the inactive interpad region while preserving timing performance (\cref{fig Ti-LGAD scheme}).

Carbonated TI-LGAD sensors have been evaluated in test beams at the CERN SPS~\cite{CERN_SPS} using \SI{120}{\giga\electronvolt} pions, evaluating timing resolution, efficiency, and interpad distance for single trench designs with varying trench widths and irradiation levels up to \fluenceUnits{2.5e15}. The irradiation of the samples was done at the TRIGA reactor\footnote{Jožef Stefan Institute, Ljubliana, Slovenia}~\cite{snoj2012computational, ambrovzivc2017computational}.

\section{Experimental Setup}
\subsection{Samples Description}

FBK (Fondazione Bruno Kessler) produced the samples within the AIDAInnova WP6 project, implementing carbon co-implantation in the gain layer to improve radiation hardness. Although various trench depths, widths, and processes were explored, this work focuses on detectors fabricated with the D2 trench-depth option, P2 process variant and V2 spacing between the gain layer and the trench, according to the manufacturer's nomenclature. the detectors have an active thickness of \SI{45}{\micro\m}~\cite{FBK-TiLGAD}. 

\begin{figure}[htbp]
    \centering
    \begin{minipage}{0.5\textwidth}
        \centering
        \begin{tabular}{l|ll}
            \hline
            \textbf{Structure} & \textbf{$\phi$/\SI{}{\fluenceUnits{}}} & \textbf{Voltage/\SI{}{\V}} \\ \hline
            TW2-TW3 & 0 & 165 \\
            TW4-TW6 & 0 & 120 \\
            TW1-TW3 & $\SI{1.5e15}{}$ & 495 \\
            TW2-TW3 & $\SI{1.5e15}{}$ & 500 \\
            TW4-TW6 & $\SI{1.5e15}{}$ & 480 \\
            TW1-TW3 & $\SI{2.5e15}{}$ & 450 \\
            TW4-TW6 & $\SI{2.5e15}{}$ & 450 \\ \hline
        \end{tabular}
        \captionof{table}{Samples tested during the testbeam including the measured bias voltages. All from Wafer 2 of the AIDAINNOVA production~\cite{FBK-TiLGAD}.}
        \label{tab samples}
    \end{minipage}
    \hfill
    \begin{minipage}{0.35\textwidth}
        \centering
        \includegraphics[width=0.6\linewidth]{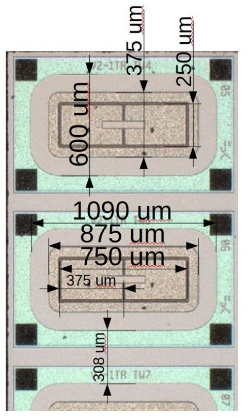}
        \caption{Dimensions of the studied structures.}
        \label{fig Ti-LGAD size}
    \end{minipage}
\end{figure}

The \cref{tab samples} lists all tested samples and their irradiation fluences. As shown in \cref{fig Ti-LGAD size}, each structure contains four readout pixels arranged such that each pair shares a single trench with varying widths (TW). TW1 corresponds to the thinnest trench, while TW6 corresponds to the thickest. Individual pixel dimensions are \SI{250}{\micro\m} \texttimes{} \SI{375}{\micro\m}.

\subsection{Test Beam Setup}\label{subsec Test beam Setup}

\begin{figure}[htbp]
    \centering
    \begin{subfigure}{0.56\linewidth}
        \centering
        \includegraphics[width=0.9\linewidth]{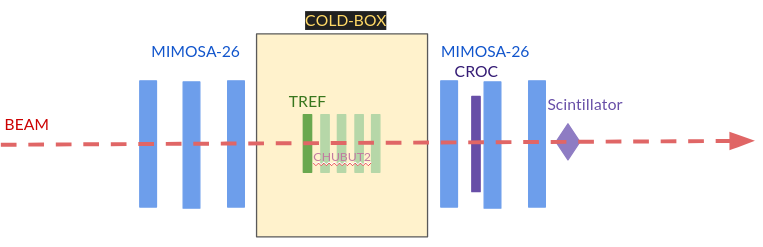}
        \caption{Sketch of the used setup.}
        \label{fig: sketch setup}
    \end{subfigure}
    \hfill
    \begin{subfigure}{0.4\linewidth}
        \centering
        \includegraphics[width=0.85\linewidth]{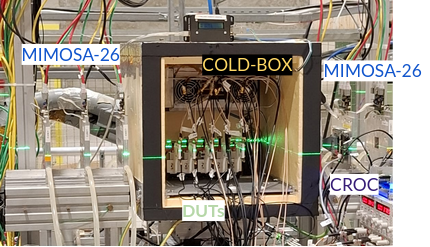}
        \caption{Picture of the used setup.}
        \label{fig: foto setup}
    \end{subfigure}
    \caption{In the sketch and the picture, the EUDET telescope made of the MIMOSA26 planes, the CROC and the coldbox with the DUTs inside are shown.}
    \label{fig: setup}
\end{figure}

Measurements were performed at the CERN SPS H6 beamline using an EUDET beam telescope~\cite{zarnecki2007eudet, jansen2016eudet} with up to six MIMOSA26 planes (\SI{18.4}{\micro\m} pitch)~\cite{baudot2009mimosa26}. The setup included a scintillator, a Trigger Logic Unit (TLU)~\cite{baesso2019aida}, and a CAEN DT5742 digitizer (500 MHz, 5 GS/s)~\cite{caen_dt5742} for DUT readout. A CROC (CMS inner tracker ReadOut Chip)~\cite{CROC} 3D module (\SI{25}{\micro\m} \texttimes{} \SI{100}{\micro\m} pitch) provided Region of Interest (ROI) selection, while an HPK LGAD served as the time reference. Up to four DUTs were tested simultaneously (\cref{fig: setup}).

The DUTs were placed inside a coldbox operated at \SI{-25}{\celsius}. They were wire-bonded to Chubut2 readout boards~\cite{chubut_testing_docs} and mounted on piezoelectric actuators~\cite{gkougkousis2025considerations} via 3D-printed holders (\cref{fig: foto setup}). This precise mechanical integration enabled micrometric alignment for triple-coincidence measurements.


The trigger is defined by a CROC-scintillator coincidende. Since the DUT readout lacks zero suppression, the digitiser records waveforms for all triggered event, regardless of whether a particle signal is present in the DUT pixel or not,  an offline analysis is required to discriminate true particle hits from empty or noise-only waveforms.

\section{Signal selection}\label{sec: waveform processing}

\begin{figure}[htbp]
    \centering
    \includegraphics[width=0.45\linewidth]{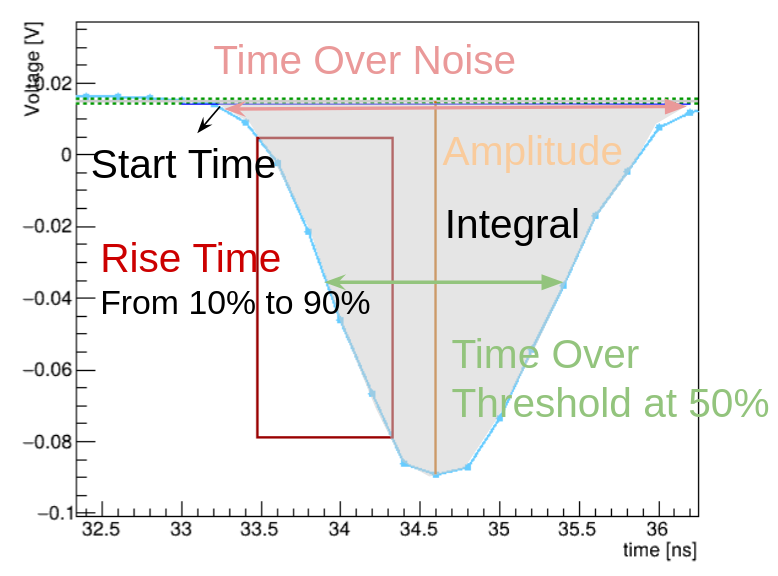}
    \caption{Examples of the waveform variables obtained from each recorded waveform.}
    \label{fig: waveform variables}
\end{figure}

For each waveform, the signal start time (last sample below baseline noise), baseline, noise, amplitude, rise time (10\%--90\%), and signal-to-noise ratio (S/N) are computed. Additionally, Time Over Noise (ToN) is calculated using a fixed threshold of $1.05 \times \text{noise}$, this time also defines the integration window for deposited charge. Finally, Time Over Threshold at 50\% ($\mathrm{ToT}_{50\%}$) provides an amplitude-independent quantity for consistent pulse shapes. These variables are illustrated in \cref{fig: waveform variables}.

\subsection{Hit Finding}

\begin{figure}[htbp]
    \centering
    \includegraphics[width=0.45\linewidth]{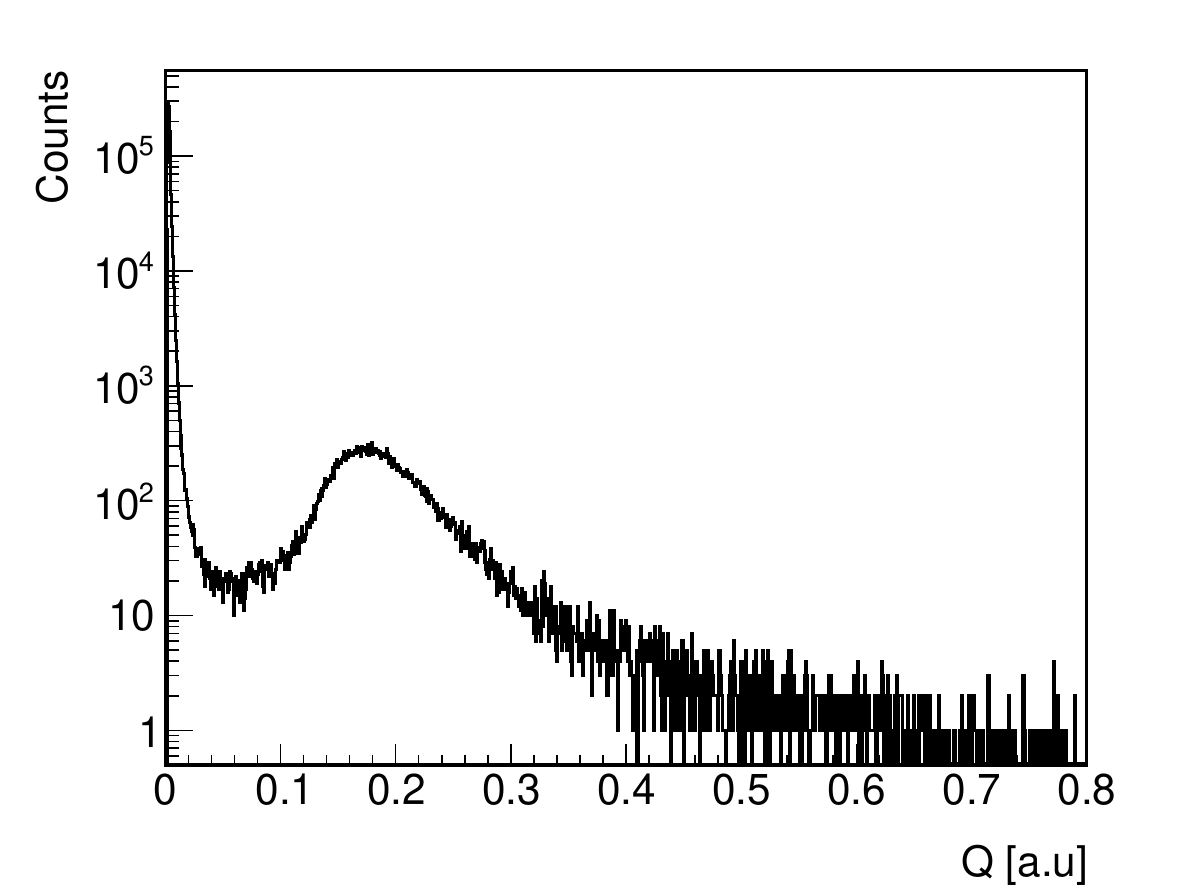}
    \caption{Charge distribution for a non irradiated TI-LGAD operated at $V=\SI{165}{V}$. Two families are observed one peaking at $Q\approx\SI{0}{\arbunit}$ and other pone peaking at $Q\approx\SI{0.2}{\arbunit}$.}
    \label{fig: charge before cuts}
\end{figure}

The procedure is going to be illustrated using  non-irradiated TI-LGAD at $V = \SI{165}{V}$. The charge deposition distribution already reveals two populations: empty waveforms ($Q\approx \SI{0}{\arbunit}$) and signal-like waveforms (peaking at $Q\approx\SI{0.2}{\arbunit}$) (\cref{fig: charge before cuts}). However, a charge-only selection is not sufficiently robust, especially when comparing devices operated at different voltages and irradiation fluences. Then, separation involves two phases: Quality Cuts followed by Selection Cuts.

\subsubsection{Quality Cuts}

\begin{figure}[htbp]
    \centering
    \begin{subfigure}{0.48\linewidth}
        \centering
        \includegraphics[width=0.9\linewidth]{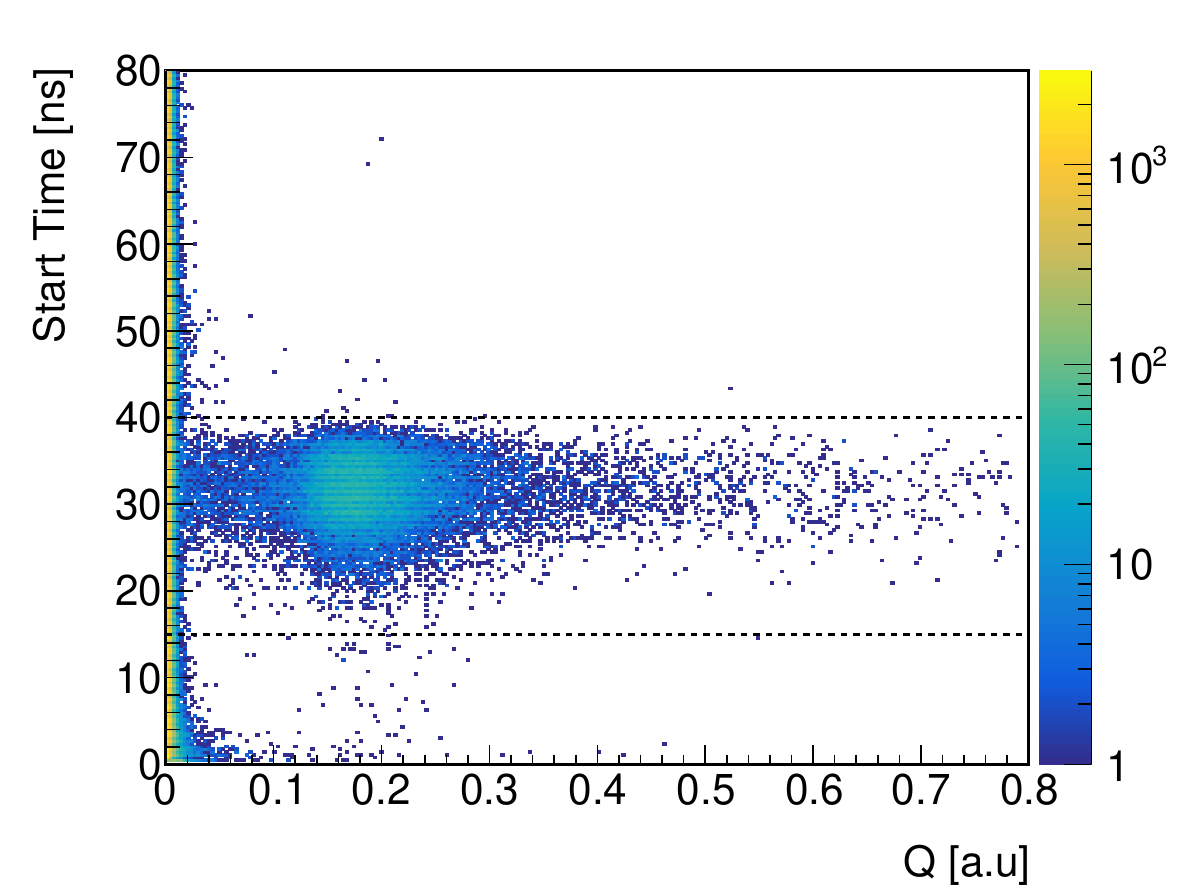}
        \caption{Start Time vs Charge. Hits with a non depreciable collected charge are confined in the Start Time region between \SI{15}{\ns} and \SI{40}{\ns}.}
        \label{fig: start time vs charge}
    \end{subfigure}
    \hfill
    \begin{subfigure}{0.48\linewidth}
        \centering
        \includegraphics[width=0.9\linewidth]{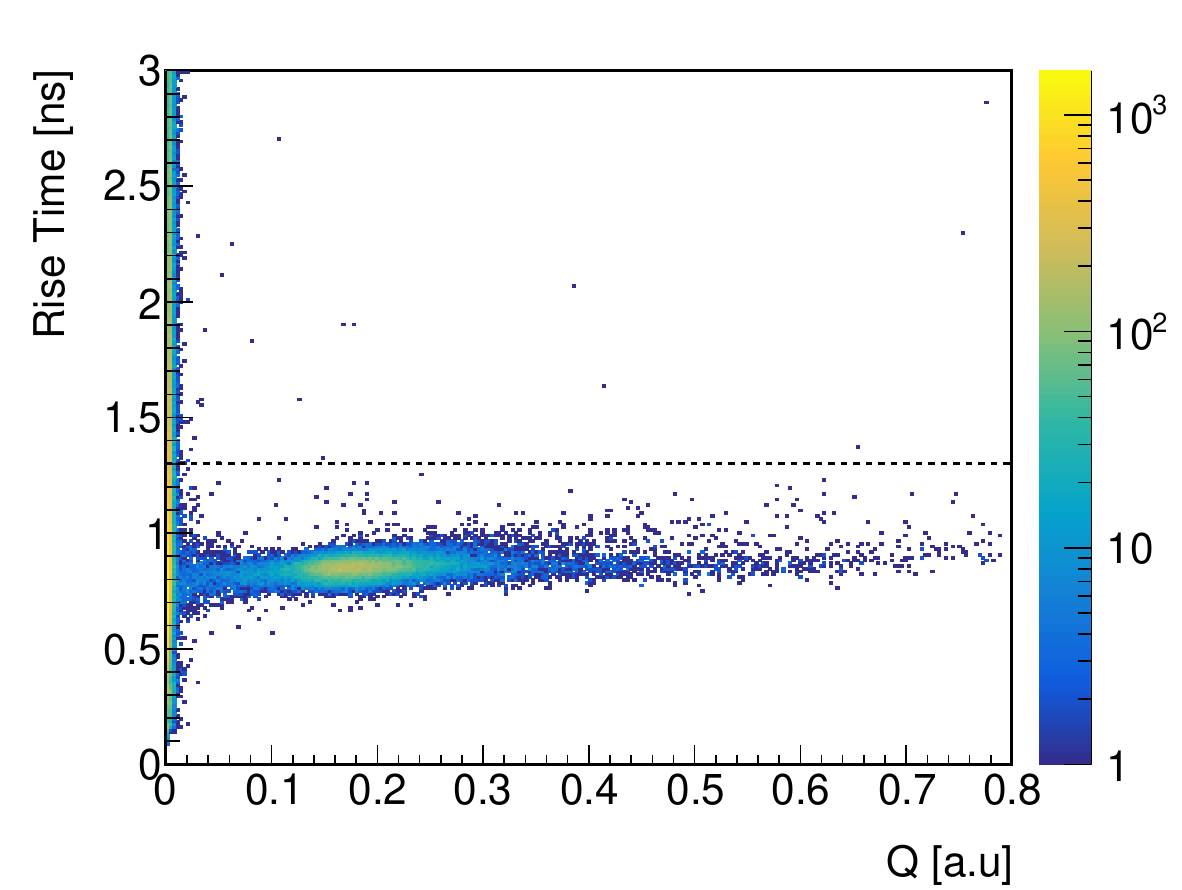}
        \caption{Rise Time vs Charge. Hits with a non depreciable collected charge have a Rise Time smaller than \SI{1.3}{\ns}.}
        \label{fig: rise time vs charge}
    \end{subfigure}
    \caption{Distributions of Start Time and Rise Time versus Charge for a non-irradiated TI-LGAD operated at $V=\SI{165}{V}$.}
    \label{fig: quality cuts}
\end{figure}

Quality cuts are first applied to remove out-of-time or non-physical waveforms. Genuine DUT signals are expected within a limited time window fixed by the trigger path. In the example shown, signals with significant charge are concentrated between \SI{15}{\ns} and \SI{40}{\ns} window (\cref{fig: start time vs charge}), and waveforms outside this interval are rejected.

A rise-time requirement is also applied. Fast LGAD pulses populate a narrow rise-time region, while slow fluctuations and low-frequency noise produce broader apparent signals. Waveforms with rise time larger than \SI{1.3}{\ns} are rejected in this example (\cref{fig: rise time vs charge}). These cuts define a clean sample of prompt, pulse-like waveforms, but do not by themselves define the final hit selection.

\subsubsection{Selection Cuts}

\begin{figure}[htbp]
    \centering
    \begin{subfigure}{0.48\linewidth}
        \centering
        \includegraphics[width=0.9\linewidth]{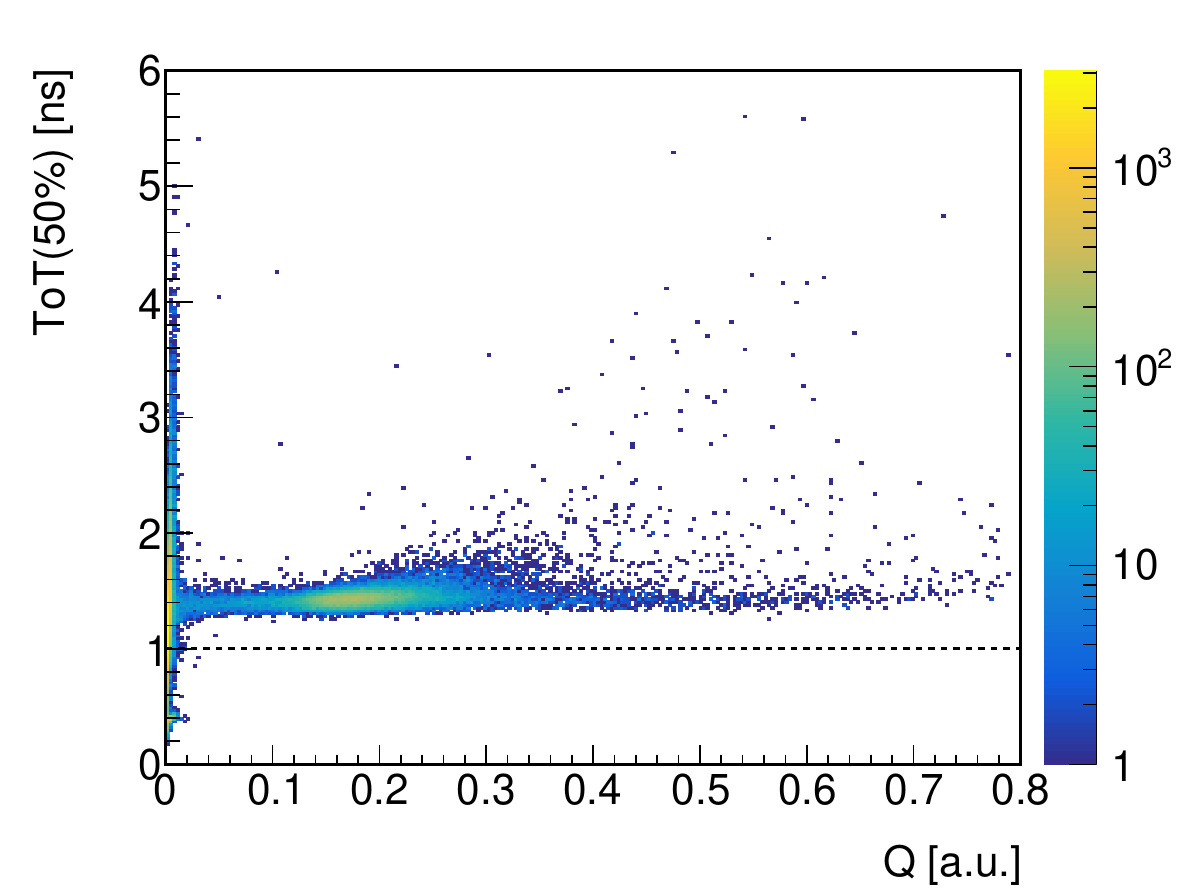}
        \caption{Distribution of the $\mathrm{ToT}_{50\%}$ versus the Charge where only hits with a Start Time between \SI{15}{\ns} and \SI{40}{\ns} and Rise Time$<\SI{1.3}{\ns}$ are being plotted. The $\mathrm{ToT}_{50\%}$ corresponding with real hits, is accumulated in the region with a $\mathrm{ToT}_{50\%}$ bigger than \SI{1}{\ns}.}
        \label{fig: tot percentage vs charge}
    \end{subfigure}
    \hfill
    \begin{subfigure}{0.48\linewidth}
        \centering
        \includegraphics[width=0.9\linewidth]{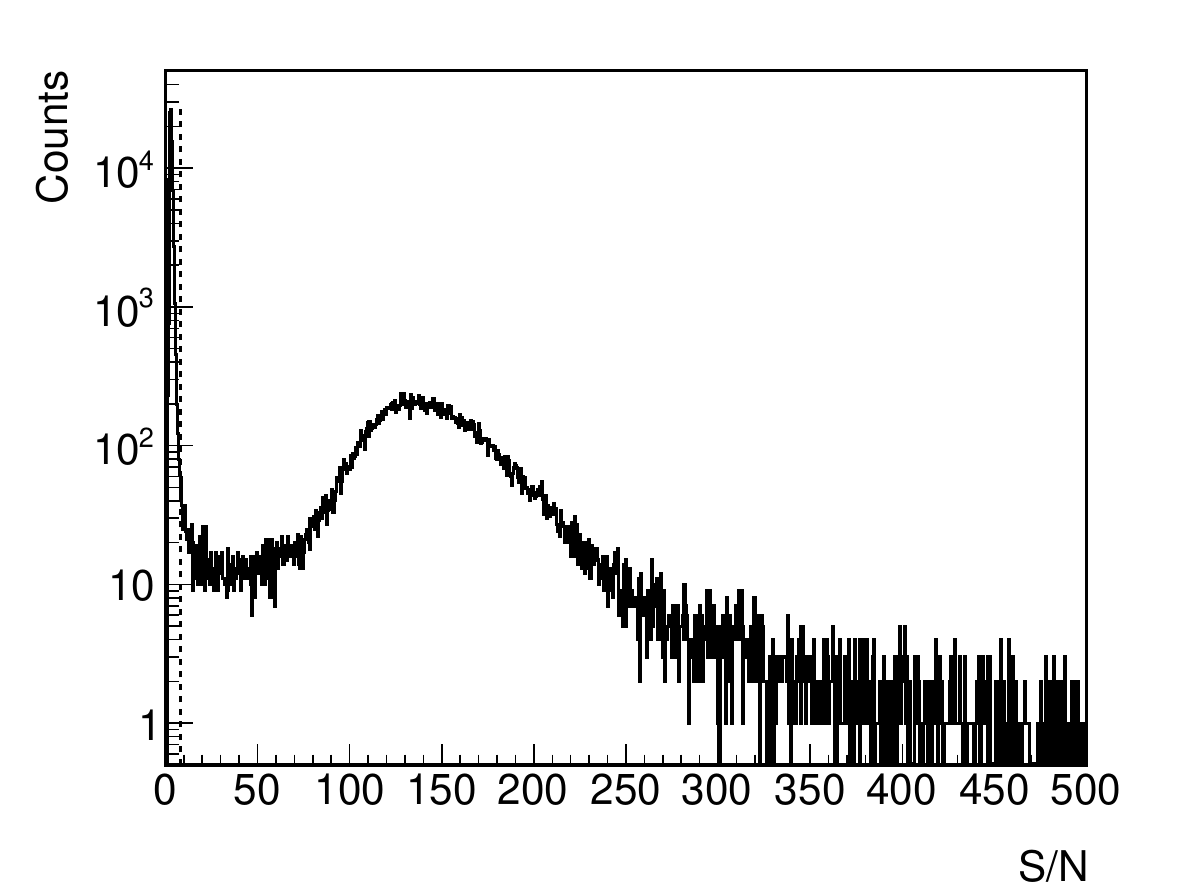}
        \caption{Distribution of the S/N where only hits with a Start Time between \SI{15}{\ns} and \SI{40}{\ns}, Rise Time$<\SI{1.3}{\ns}$ and $\text{ToT}_\%\geq\SI{1}{\ns}$ are being plotted. There are two families, one peaking at S/N$\approx 2$, the noise, and the other peaking at S/N$\approx 150$, the real hits.}
        \label{fig: snr}
    \end{subfigure}
    \caption{Distributions of $\mathrm{ToT}_{50\%}$ and signal noise ratio for a non-irradiated TI-LGAD operated at $V=\SI{165}{V}$.}
    \label{fig: selection cuts}
\end{figure}

The final hit selection uses ($\mathrm{ToT}_{50\%}$) and signal-to-noise ratio. The ($\mathrm{ToT}_{50\%}$) requirement rejects short noise fluctuations and keeps waveforms compatible with the expected LGAD pulse shape. In the example device, genuine signals populate the region $\mathrm{ToT}_{50\%}>\SI{1}{\ns}$ (\cref{fig: tot percentage vs charge}).

After these requirements, the $S/N$ distribution shows a clear separation between empty waveforms and real DUT signals. The $S/N$ threshold is chosen by balancing hit efficiency and fake-hit rate. For the example shown, a threshold $S/N>8$ is used (\cref{fig: snr}), corresponding to a fake-hit contribution below the required level.

The same strategy is applied to all measured devices. The numerical values of the thresholds may depend on the sensor, bias voltage, and irradiation fluence, but the logic of the selection is common to the full dataset.




\subsection{Analysis Observables and Reconstruction Methodology}\label{sec: Analysis Variables}

Testbeam data was analyzed using the Corryvreckan framework~\cite{dannheim2021corryvreckan}. Tracks reconstructed from MIMOSA26 hits were filtered to retain only those matched to a CROC hit (``in-time'' tracks), suppressing out-of-time contributions coming from the MIMOSA integration time ($\mathcal{O}(200\,\si{\micro\second})$).

Detector efficiency ($\epsilon$) is defined as:
\begin{equation}
    \epsilon = \frac{\text{In-time tracks}\cap\text{Hit in the DUT}}{\text{In-time tracks}}
    \label{eq eficiencia}
\end{equation}

The efficiency profile across the sensor (including edges and trenches) is characterised as a double error function to model the step-like response smeared by the finite Gaussian spatial resolution:
\begin{equation}
f(x) =
\text{Plateau}\cdot \left[
\operatorname{erf}\!\left(\frac{x - \mu_1}{\sigma_1 \sqrt{2}}\right)
-
\operatorname{erf}\!\left(\frac{x - \mu_2}{\sigma_2 \sqrt{2}}\right)
\right]
 + \text{Offset.}
 \label{eq: double error function}
\end{equation}
Although two independent width parameters, $\sigma_1$ and $\sigma_2$, are introduced to improve the stability of the fit, the underlying spatial resolution is assumed to be described by a single Gaussian width. An effective resolution is therefore extracted by combining both parameters, taking the mean value $\bar{\sigma}$ and assigning an uncertainty given by their spread.

The finite spatial resolution is estimated from local residuals using a clean sample of isolated in-time tracks in the closest telescope plane to the DUTs, the third one. It was obtained a resolution of $\sim\SI{3.85\pm0.04}{\um}$ (six telescope planes configuration) and $\sim\SI{4.0\pm0.2}{\um}$ (five planes configuration).
    
    

Timing performance was evaluated with a three-sensor coincidence method, using a Constant Fraction Discriminator (CFD) to reduce the amplitude-dependent time walk~\cite{CFD}. The time resolution $\sigma_i$ for each DUT is derived from the Gaussian widths of pairwise time differences:
\begin{equation}
\sigma_i = \sqrt{\frac{\sigma^2_{i-j} + \sigma^2_{i-k} - \sigma^2_{j-k}}{2}}, \quad (i,j,k)\ \text{cyclic permutations of }(1,2,3).
\label{eq time resolution}
\end{equation}
Where $\sigma_{a-b}$ denotes the standard deviation of the ToA difference between detectors $a$ and $b$, obtained from a Gaussian fit. The corresponding uncertainties $\delta_i$ are computed as function of the errors $\delta_{1-2}$, $\delta_{1-3}$ and $\delta_{2-3}$:
\begin{equation}
    \delta_i = \frac{\sqrt{(\sigma_{1-2}\delta_{1-2})^2+(\sigma_{1-3}\delta_{1-3})^2 + (\sigma_{2-3}\delta_{2-3})^2}}{2\sigma_i} \quad i = 1, 2, 3 \text{.}
\label{eq error time resolution}
\end{equation}

\section{Results}

\subsection{Tracking Results}

\begin{figure}[htbp]
    \centering

    \begin{subfigure}[t]{0.5\textwidth}
        \centering
        \includegraphics[width=0.8\linewidth]{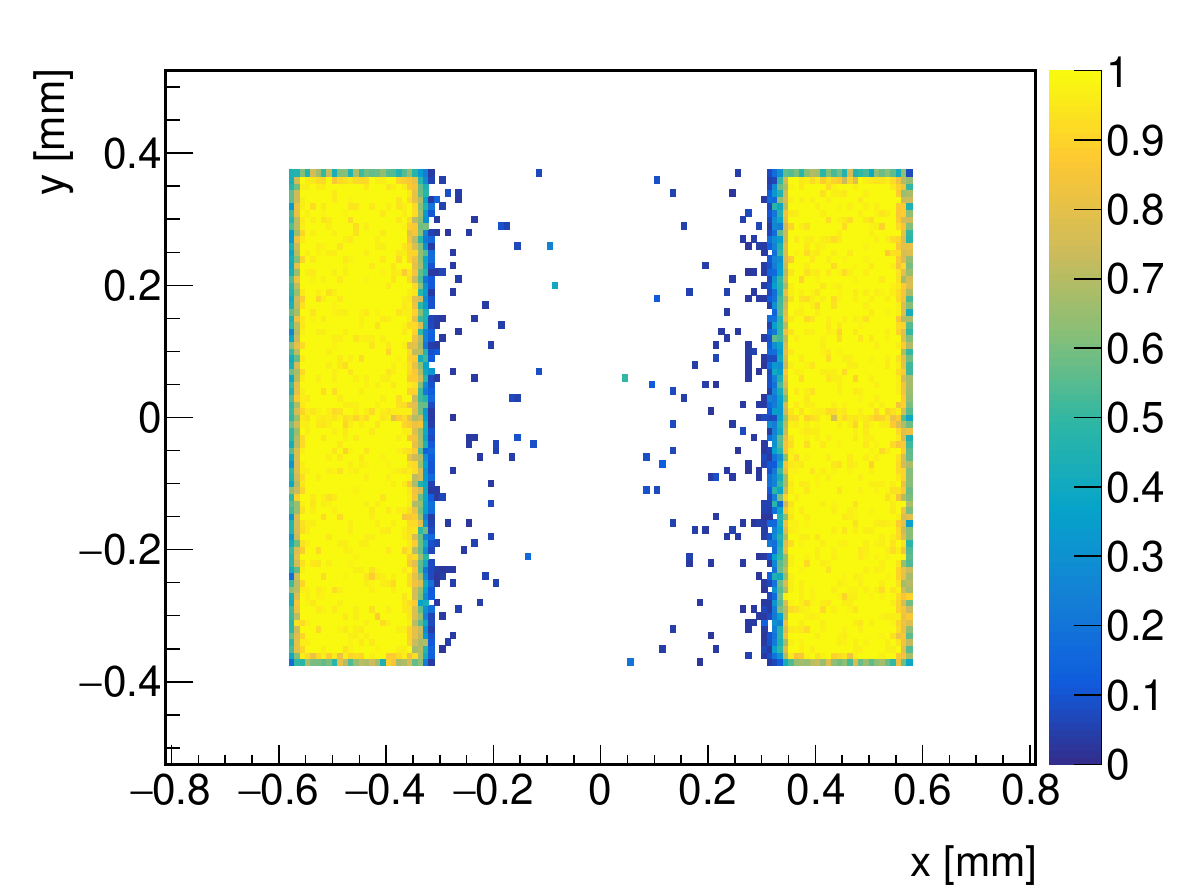}
        \caption{2D efficiency plot. Left structure is TW2 and right one is TW3}
        \label{fig: Batch3 DUT1 Efficiency: 2D}
    \end{subfigure}

    \begin{subfigure}[t]{0.45\textwidth}
        \centering
        \begin{tikzpicture}
            \node[anchor=south west, inner sep=0] (main) at (0,0) {
                \includegraphics[width=0.8\linewidth]{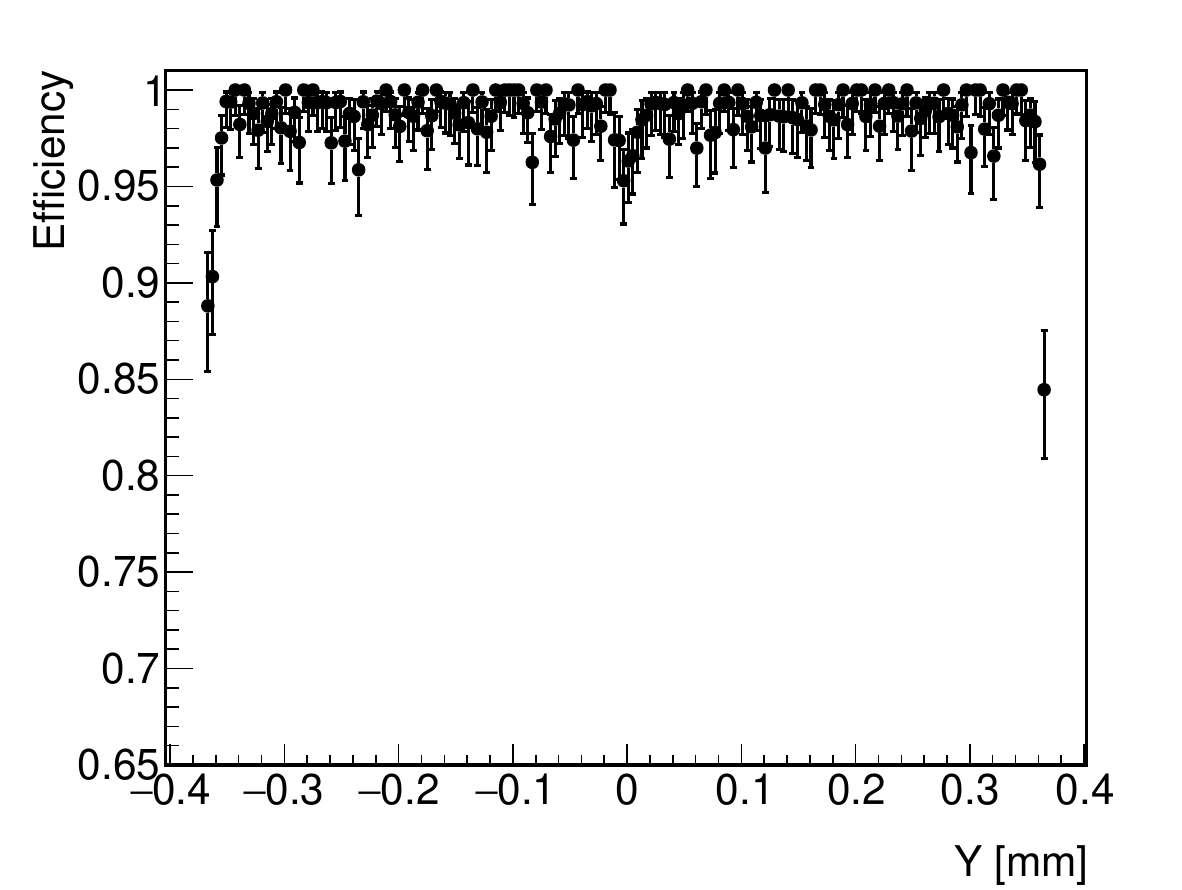}
            };

            \node[
                anchor=center,
                fill=white,
                opacity=0.7,
                inner sep=2pt
            ] at ($(main.center)+(0.1,-0.3cm)$) {
                \includegraphics[width=0.4\linewidth]{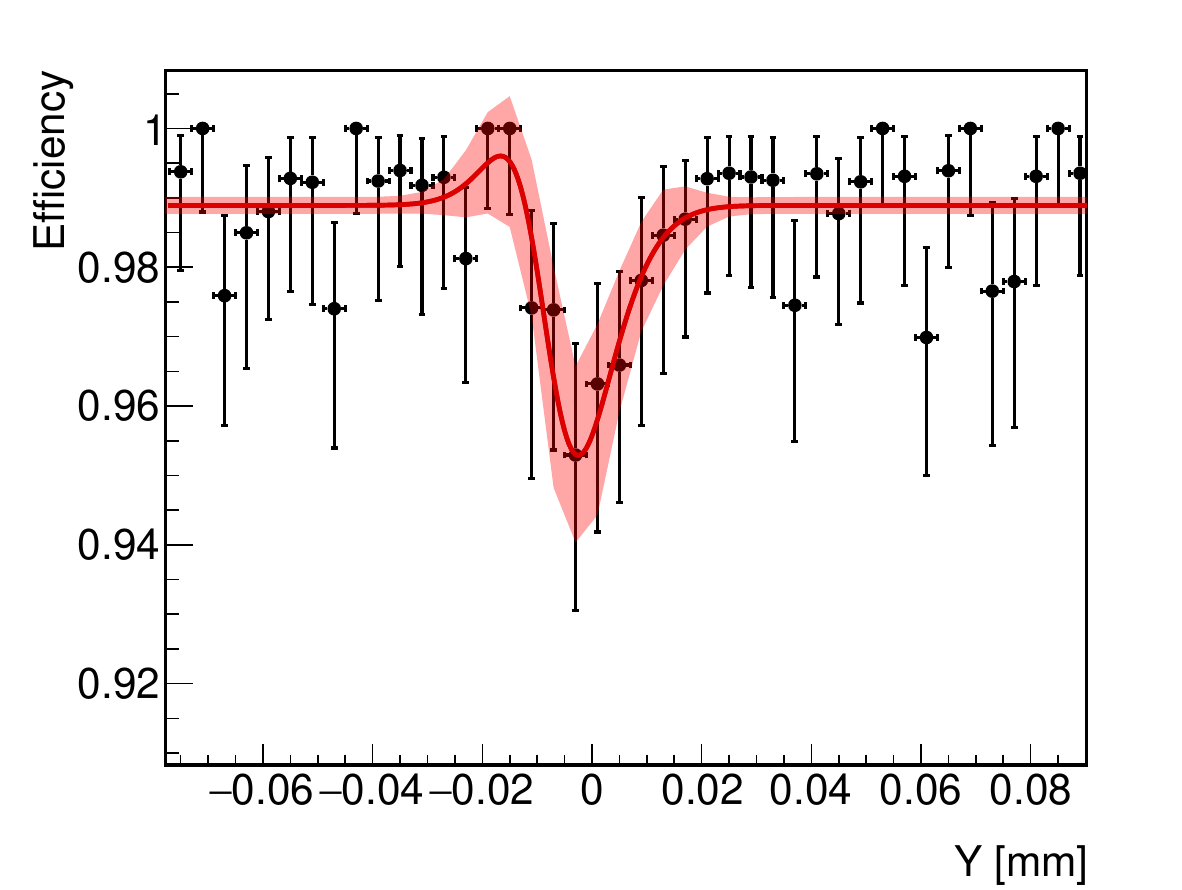}
            };
        \end{tikzpicture}
        \caption{TW2 structure efficiency projection. The red line shows the best fit to a double error function (\cref{eq: double error function}) with parameters: $\text{Plateau}=\SI{-0.7\pm0.5}{}$, $\mu_1=\SI{-0.005\pm0.006}{\mm}$, $\sigma_1=\SI{0.009\pm0.005}{\mm}$, $\mu_2=\SI{-0.004\pm0.009}{\mm}$, $\sigma_2=\SI{0.009\pm0.005}{\mm}$ and $\text{Offset}=\SI{0.990\pm0.001}{}$.}
        \label{fig: Batch3 DUT1 Efficiency: Projection TW2}
    \end{subfigure}
    \hfill
    \begin{subfigure}[t]{0.45\textwidth}
        \centering
        \begin{tikzpicture}
            \node[anchor=south west, inner sep=0] (main) at (0,0) {
                \includegraphics[width=0.8\linewidth]{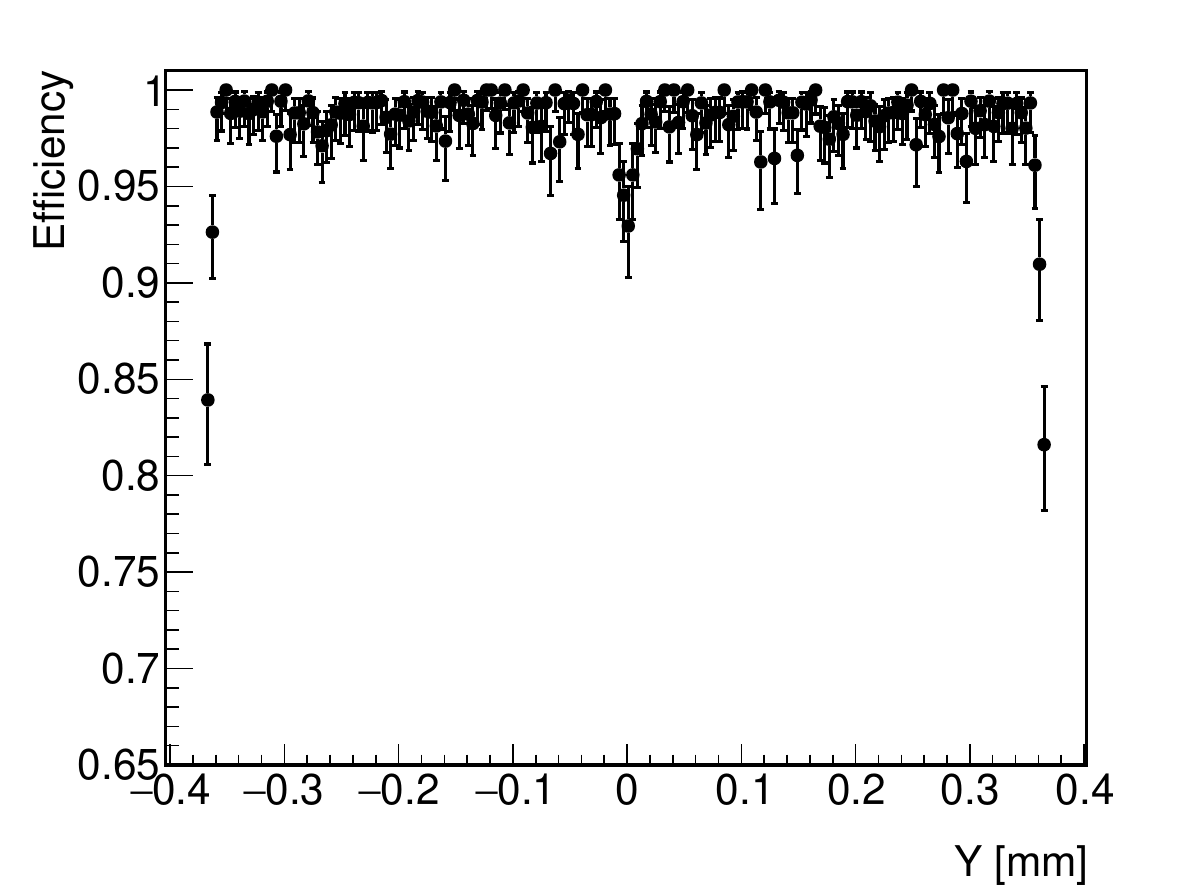}
            };

            \node[
                anchor=center,
                fill=white,
                opacity=0.7,
                inner sep=2pt
            ] at ($(main.center)+(0.1,-0.3cm)$) {
                \includegraphics[width=0.4\linewidth]{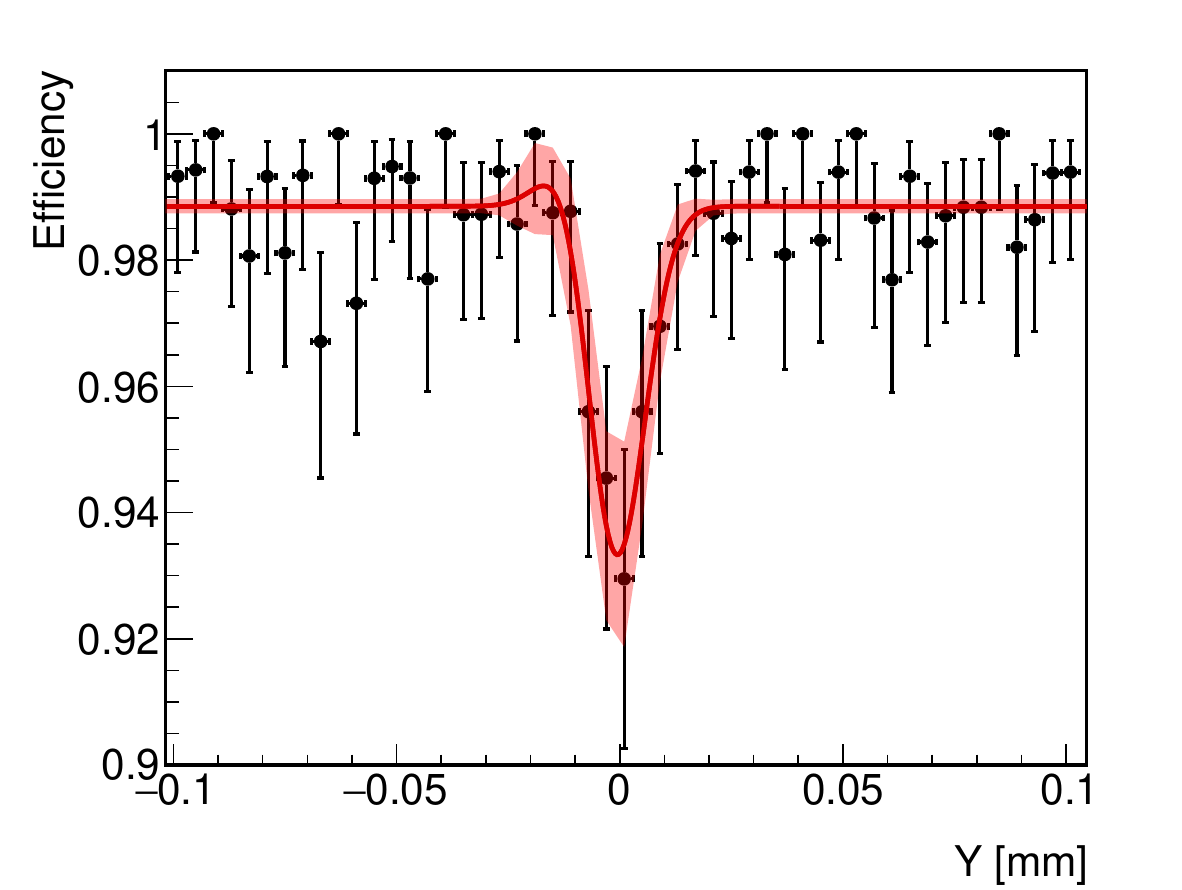}
            };
        \end{tikzpicture}
        \caption{TW3 structure efficiency projection. The red line shows the best fit to a double error function (\cref{eq: double error function}) with parameters: $\text{Plateau}=\SI{-0.03\pm0.04}{}$, $\mu_1=\SI{-0.008\pm0.04}{\mm}$, $\sigma_1=\SI{0.002\pm0.05}{\mm}$, $\mu_2=\SI{0.006\pm0.015}{\mm}$, $\sigma_2=\SI{0.008\pm0.009}{\mm}$ and $\text{Offset}=\SI{0.99\pm0.05}{}$.}
        \label{fig: Batch3 DUT1 Efficiency: Projection TW3}
    \end{subfigure}
    \caption{Measured hit efficiency for  non-irradiated TI-LGADs with TW2 and TW3
operated at $V = \SI{165}{\V}$. }
    \label{fig: Batch3 DUT1 Efficiency}

\end{figure}

An efficiency map is obtained for each detector by matching ``in-time'' tracks to DUT hits (\cref{fig: Batch3 DUT1 Efficiency: 2D}). Representative examples of projections perpendicular to the trench are shown in \cref{fig: Batch3 DUT1 Efficiency: Projection TW2,fig: Batch3 DUT1 Efficiency: Projection TW3}. They show that the efficiency in the trench region remains above 94\% for TW2 and 90\% for TW3. The interpad distance is extracted by fitting the efficiency profile to a double error function and solving for a fixed efficiency value, with errors estimated from the $1\sigma$ confidence interval of the fit.

\begin{figure}[htbp]
    \centering
    \begin{subfigure}{0.48\linewidth}
        \centering
        \includegraphics[width=0.85\linewidth]{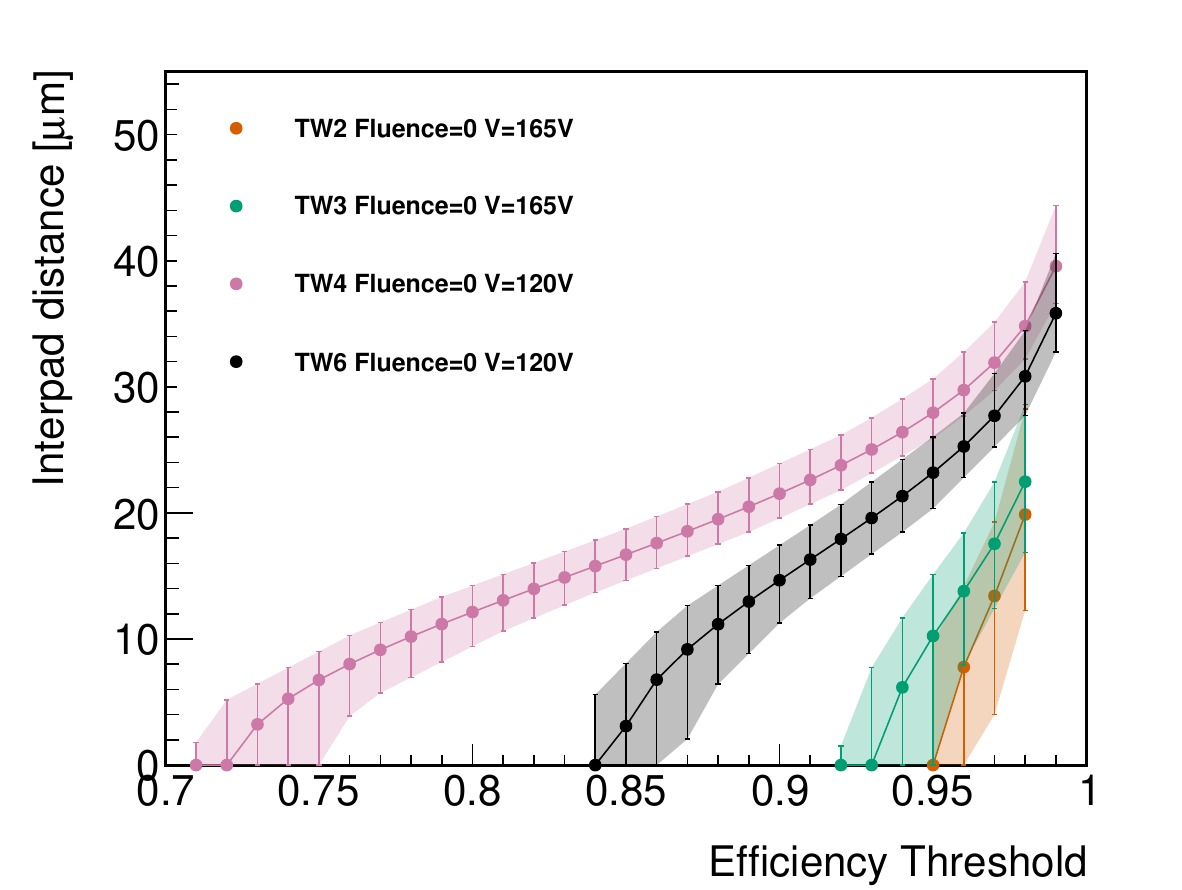}
        \caption{Non irradiated devices.}
        \label{fig: interpad vs efficiency: no irrad}

    \end{subfigure}
    \hfill
    \begin{subfigure}{0.48\linewidth}
        \centering
        \includegraphics[width=0.85\linewidth]{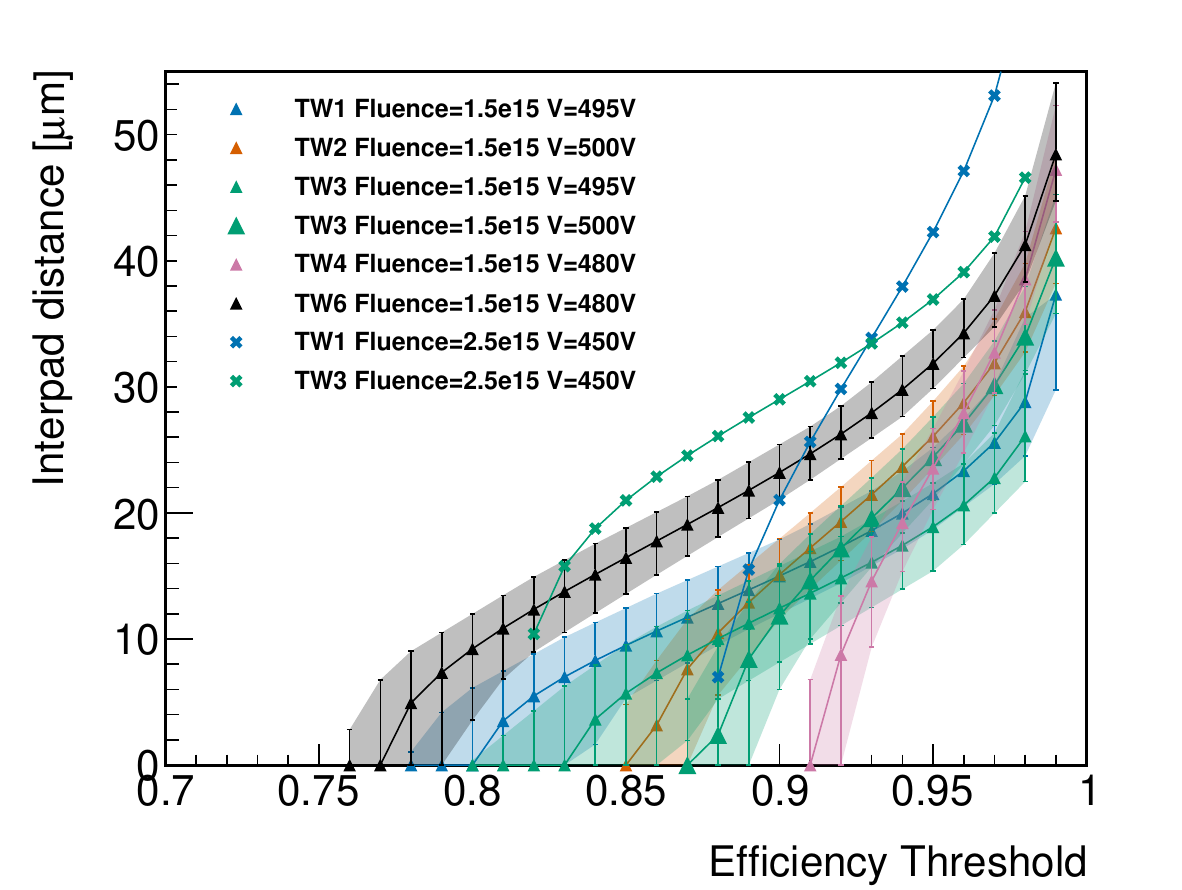}
        \caption{Irradiated devices.}
        \label{fig: interpad vs efficiency: irrad}

    \end{subfigure}
    
    \caption{Comparison of the interpad distance for different requested efficiencies between all measured detectors.}
    \label{fig: interpad vs efficiency}
\end{figure}

The inter-pad distance is compared for all measured devices listed in \cref{tab samples} in the \cref{fig: interpad vs efficiency}. Larger inter-pad distances are observed for the highly irradiated devices, indicating a degradation of the efficiency in the trench region after irradiation. For the non-irradiated devices (\cref{fig: interpad vs efficiency: no irrad}), TW2 shows the smallest inter-pad loss, while TW3 exhibits a larger loss at high efficiency thresholds.

For the irradiated devices, the effective inter-pad distance increases with fluence, with the largest values observed for the samples irradiated to $\SI{2.5e15}{\fluenceUnits{}}$. At $\SI{1.5e15}{\fluenceUnits{}}$, TW4 shows the smallest inter-pad distance at moderate efficiency thresholds, while TW3 operated at $\SI{500}{\volt}$ performs better at the highest requested efficiencies. The comparison also shows a sizeable dependence on the applied bias voltage, as illustrated by the two TW3 measurements at $\SI{495}{\volt}$ and $\SI{500}{\volt}$. Overall, TW2--TW4 provide the best compromise, depending on fluence, bias voltage, and requested efficiency.

\subsection{Charge collection with Tracking}

\begin{figure}[htbp]
    \centering
    \begin{subfigure}{0.5\textwidth}
        \centering
        \includegraphics[width=0.8\linewidth]{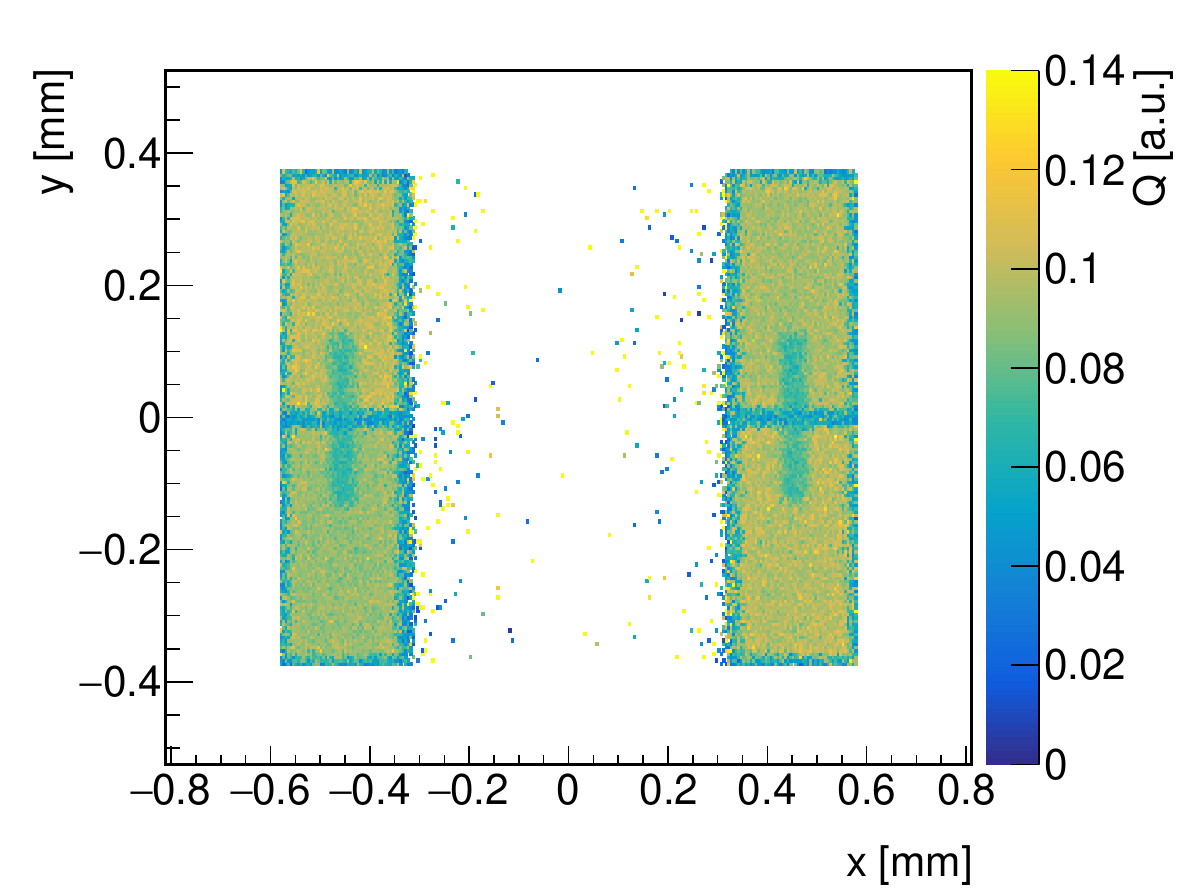}
        \caption{2D map. Left structure is TW2 and right one is TW3.}
        \label{fig: Batch2 DUT2 Charge: 2D}
    \end{subfigure}
    
    \begin{subfigure}{0.45\textwidth}
        \centering
        \includegraphics[width=0.8\linewidth]{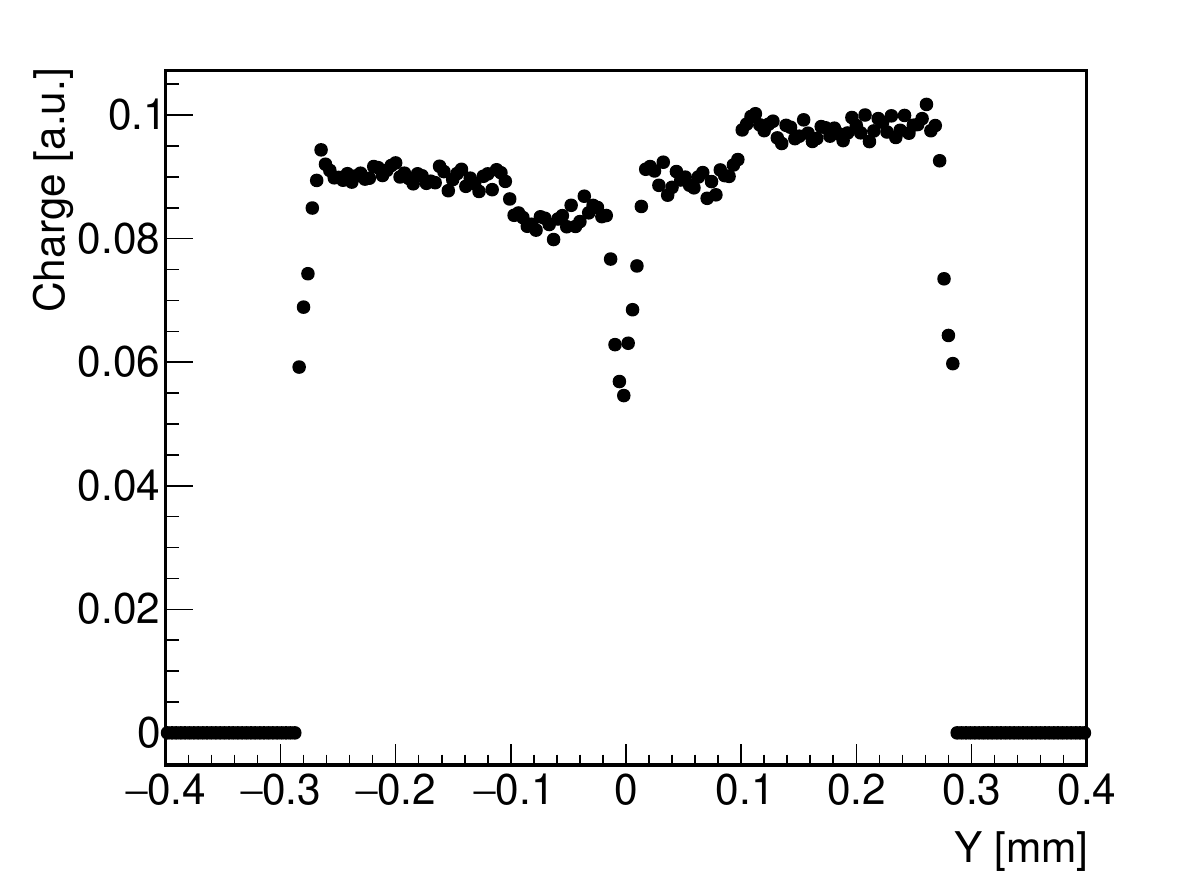}
        \caption{TW2 mean of collected charge MPV projection.}
        \label{fig: Batch2 DUT2 Charge: Projection TW2}
    \end{subfigure}
    \begin{subfigure}{0.45\textwidth}
        \centering
        \includegraphics[width=0.8\linewidth]{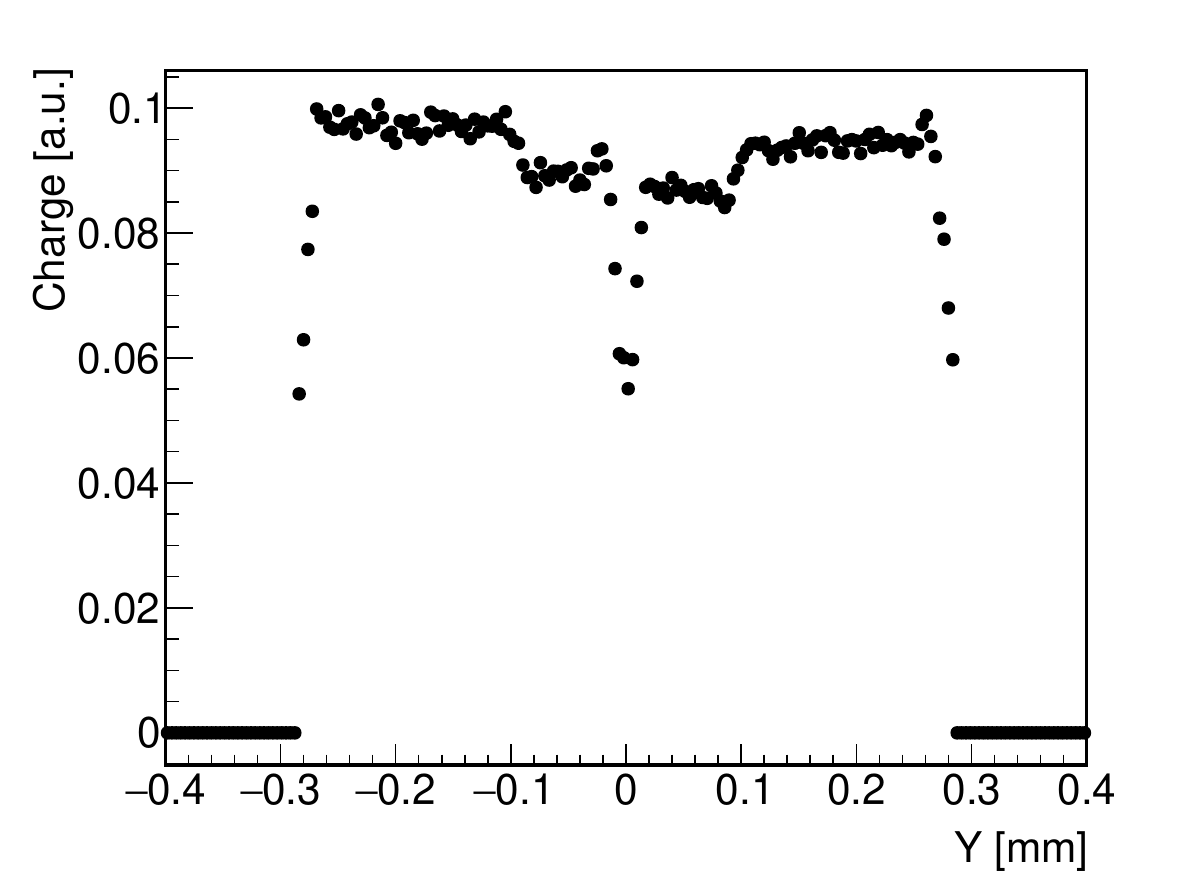}
        \caption{TW3 mean of collected charge MPV projection.}
        \label{fig: Batch2 DUT2 Charge: Projection TW3}
    \end{subfigure}
    \caption{Collected charge MPV results for a irradiated ($\phi = \SI{1.5e15}{\fluenceUnits{}}$) TI-LGADs with TW2 and TW3 operated at $V = \SI{500}{\V}$.}
    \label{fig: Batch2 DUT2 Charge}
\end{figure}

A 2D map of the most probable collected charge per bin was obtained using track position information (\cref{fig: Batch2 DUT2 Charge: 2D}). Projections perpendicular to the trench (\cref{fig: Batch2 DUT2 Charge: Projection TW2,fig: Batch2 DUT2 Charge: Projection TW3}) show around $\SI{60}{\%}$ decrease in the interpad region relative to the pixel center. A vertical region of reduced charge corresponds to the optical window used for laser measurements, this is only observed for irradiated devices (the showed device here correspond with a fluence of $\phi = \SI{1.5e15}{\fluenceUnits{}}$). The step between pixels is attributed to a loose contact in the Chubut-2 boards.

\subsection{Timing Results}

\begin{figure}[htbp]
    \centering

    \begin{subfigure}{0.3\textwidth}
        \centering
        \includegraphics[width=\linewidth]{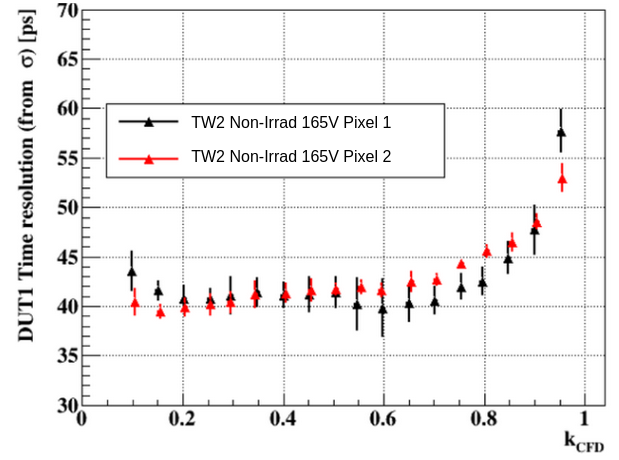}
        \caption{Non-irradiated TI-LGAD with TW2 and operated at $V=\SI{165}{\V}$}
        \label{fig: time resolution 0}
    \end{subfigure}
    \hfill
    \begin{subfigure}{0.33\textwidth}
        \centering
        \includegraphics[width=0.9\linewidth]{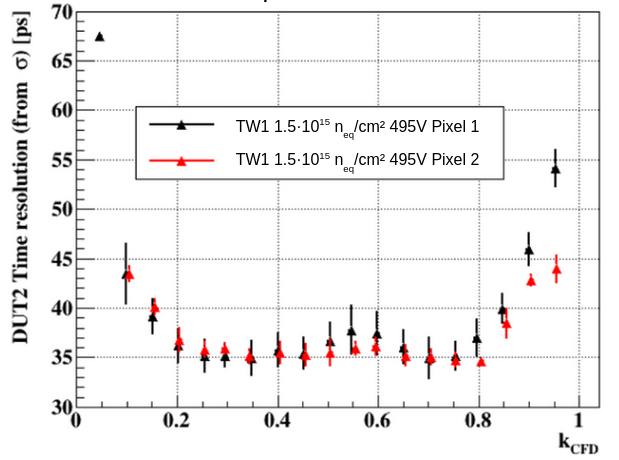}
        \caption{Irradiated TI-LGAD ($\phi = \SI{1.5e15}{\fluenceUnits{}}$) with TW1 and operated at $V=\SI{495}{\V}$}
        
        \label{fig: time resolution 1.5}
    \end{subfigure}
    \hfill
    \begin{subfigure}{0.33\textwidth}
        \centering
        \includegraphics[width=0.9\linewidth]{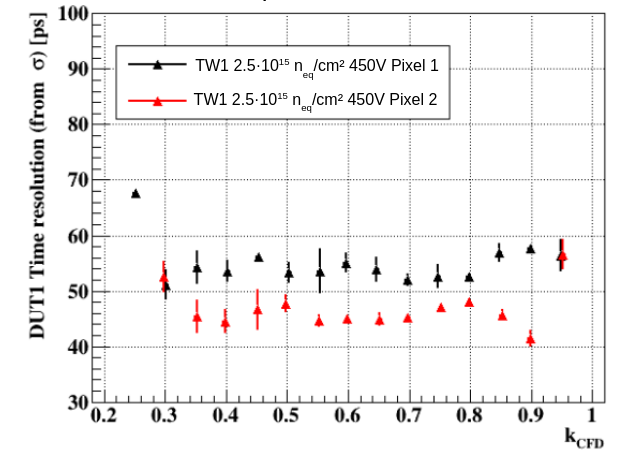}
        \caption{Irradiated TI-LGAD ($\phi = \SI{2.5e15}{\fluenceUnits{}}$) with TW1 and operated at $V=\SI{450}{\V}$}
        \label{fig:time resolution 2.5}
    \end{subfigure}

    \caption{Time resolution of three different TI-LGADs as a function of the CFD fraction.}
    \label{fig: time resolution}
\end{figure}

Timing resolution was computed using MIP signals and the three-sensor method (\cref{sec: Analysis Variables}) with a CFD used to reduce amplitude-dependent time walk. Results (\cref{fig: time resolution}) show resolutions ($\sigma_t$) of $\sim\SI{40}{\ps}$ (non-irradiated), $\sim\SI{35}{\ps}$ (for devices irradiated to $\phi=\SI{1.5e15}{\fluenceUnits{}}$), and $\sim\SI{50}{\ps}$ (for devices irradiated to $\phi=\SI{2.5e15}{\fluenceUnits{}}$).

The TW1 device irradiated to $\phi=\SI{1.5e15}{\fluenceUnits{}}$ achieved the best resolution among the three devices shown. This improvement over the unirradiated case is attributed to a faster rise time from the higher bias voltage (stronger electric field), while the reduced gain after irradiation may also mitigate the impact of Landau fluctuations.


\section{Conclusions}

A characterization of neutron-irradiated (up to  $\SI{2.5e15}{\fluenceUnits{}}$) carbon-enriched V2 TI-LGAD sensors with different trench widths has been presented. The performance of the TI-LGAD concept has been experimentally evaluated using minimum ionizing particles.

For the best-performing geometries, the effective interpad distance is compatible with zero for efficiency thresholds of $\ge 95\%$ before irradiation and $\ge 90\%$ after irradiation to $\SI{1.5e15}{\fluenceUnits{}}$, with full efficiency recovered outside the inter-pad region. Time resolutions between approximately $\SI{35}{\ps}$ and $\SI{50}{\ps}$ are obtained for the studied fluences (up to $\SI{2.5e15}{\fluenceUnits{}}$), with the best value measured for the device irradiated to $\SI{1.5e15}{\fluenceUnits{}}$. The optimal geometries are identified as V2 TW2 D2 P2 before irradiation and V2 TW4 D2 P2 after irradiation.

Future work will evaluate efficiency versus track inclination, extract detailed time-resolution maps, and optimize trench parameters for DRD3 production.





\acknowledgments
This work was supported by the Programa de Ayudas Predoctorales Concepción Arenal of the University of Cantabria, co-funded by the Government of Cantabria. It was also co-funded by the Complementary Plan in Astrophysics and High-Energy Physics (CA25944), project C17.I02.P02.S01.S03 CSIC CERN, supported by the Next Generation EU funds, RRF and PRTR mechanisms, and the Government of the Autonomous Community of Cantabria.

This work was developed within the framework of the CERN DRD3 collaboration and was funded by the Spanish Ministry of Science, Innovation and Universities (MICIU/AEI/10.13039/501100011033) and by the European Regional Development Fund (ERDF) program “A way of making Europe” under Grant PID2023-148418NB-C41. Additional support was provided by the European Union’s Horizon 2020 Research and Innovation Programme through AIDAinnova (Grant Agreement No. 101004761).

This work also received support from the European Union NextGenerationEU/PRTR project C17.I02.P02 – SGI\_GICS\_Nuevas actuaciones en grandes infraestructuras de investigación europeas e internacionales, and from Grant RyC-2023-044327-I funded by MICIU/AEI/10.13039/501100011033 and by FSE+, co-funded by the European Social Fund program ``El FSE invierte en tu futuro'', as well as grant PRE2019-087514.

This work was supported by Grant CSN2024-154303 funded by MICIU/AEI/10.13039/501100011033 


\bibliographystyle{JHEP}
\bibliography{biblio.bib}

\end{document}